# A Grain Boundary Embrittlement Genome for Substitutional Cubic Alloys

Nutth Tuchinda[a*], Gregory B. Olson[a], Christopher A. Schuh[a, b]

[a]Department of Materials Science and Engineering, Massachusetts Institute of Technology, 77 Massachusetts Avenue, Cambridge, MA, 02139, USA

[b]Department of Materials Science and Engineering, Northwestern University, Clark Street 633, Evanston, IL, 60208, USA

*Correspondence to nutthtu@mit.edu

## Abstract

Grain boundary chemistry plays a critical role for the properties of metals and alloys, yet there is a lack of consistent datasets for alloy design and development. With the advent of artificial intelligence and machine learning in materials science, open materials models and datasets can be used to overcome such challenges. Here, we use a universal interatomic potential to compute a grain boundary segregation and embrittlement genome for the Σ5[001](210) grain boundary for FCC and BCC binary alloys. The grain boundary database calculated here serves as a design tool for the embrittlement of high-angle grain boundaries for alloys across 15 base metals system of Ag, Al, Au, Cr, Cu, Fe (both BCC and FCC), Mo, Nb, Ni, Pd, Pt, Rh, Ta, V and W with 75 solute elements for each.

Grain boundary (GB) solute segregation is often associated with GB embrittlement, as for the well-known Fe(S), Fe(P) and Fe(Sn) systems[1–5]. However, many alloying elements are not embrittling, or do not segregate in the first place. Some combinations of solvent (host) and GB segregant lead to strengthening of the boundaries[3,6–10], or provide other beneficial properties such as thermal stability[11–14] and improved mechanical properties[15–17]. Successful alloy design therefore increasingly requires a nuanced understanding of GB segregation and embrittlement. The past few years have seen remarkable progress in the understanding of the segregation part of this problem, with large atlases of data being presented for the thermodynamic quantities that govern segregation in the full range of atomic sites present in GBs in polycrystalline environments[18,19]. However, the embrittlement part of this problem remains an area where large atlases of self-consistent data are not yet available for many alloys. Recent attempts to aggregate published datasets[8,20] illustrate challenges in cross-comparing amongst data generated with a variety of methodologies[8,21–23]. Moreover, the methodology of assessing GB embrittlement potency is based on GB slab methods that often require significant computational resources[24–26]. As a result, there are limited data on GB segregation and embrittlement for use in computational alloy design frameworks[27,28].

In this letter, we address this challenge using the recent advancements in artificial intelligence and machine learning[29–32], to provide a large, self-consistent analysis of GB embrittlement potency for ~1,000 binary alloys spanning a large set of relevant alloys. We leverage open materials datasets[29,33–36] and the EquiformerV2 (eqV2) model[37,38] in order to achieve great accuracy vis-à-vis density functional theory (DFT) at a fraction of the cost, as benchmarked on Matbench Discovery[39]. Specifically, we apply the small (31M parameters) model pretrained with the Open Materials dataset[34] and fine-tuned with MPtrj[35] and Alexandria[36] dataset (eqV2_31M_omat_mp_salex[33,34]) to calculate segregation tendency and embrittlement potency. We consider 15 base metals: Ag, Al, Au, Cr, Cu, Fe (both BCC and FCC), Mo, Nb, Ni, Pd, Pt, Rh, Ta, V and W, in combination with 75 solute elements each (He, Li, Be, Na, Mg, Al, Si, P, S, Cl, Ar, K, Ca, Sc, Ti, V, Cr, Mn, Fe, Co, Ni, Cu, Zn, Ga, Ge, As, Se, Br, Kr, Rb, Sr, Y, Zr, Nb, Mo, Tc, Ru, Rh, Pd, Ag, Cd,

In, Sn, Sb, Te, I, Xe, Cs, Ba, La, Ce, Pr, Nd, Pm, Sm, Eu, Gd, Tb, Dy, Ho, Er, Tm, Yb, Lu, Hf, Ta, W, Re, Os, Ir, Pt, Au, Hg, Tl, Pb and Bi).

We construct a Σ5[001](210) GB slab model[40] using GB_code[41] with a 2×2×2 repeating supercell for FCC metals, and 3×2×2 for BCC metals. The relaxations of the GBs are done with the Fast Inertial Relaxation Engine (FIRE) minimization algorithm[42,43] with a force tolerance of 0.01 eV/Å. Analyses in this work are conducted with Python software packages from Refs.[44–55]. We provide the values for the lattice constants, supercell sizes, system sizes and their respective GB energy in Table I along with a reported GB energy from the literature. We also show an example structure of the BCC Fe and FCC Al grain boundaries in Fig. 1, with all 4 sites used for segregation calculations. The GB energies computed here are reasonably matched to the DFT dataset[56,57], see Table I.

TABLE I Calculated lattice constants and their corresponding Σ5[001](210) grain boundary supercell size. The grain boundary energies from the present work are also compared with the density functional theory data taken from Ref.[56], except for BCC Fe which is from Ref.[57].

| System | Supercell Size (Å) | GB Slab Size (atoms) | GB Energy (J/m²) | DFT GB Energy (J/m²) |
|---|---|---|---|---|
| Ag | 8.308×9.288×52.153 | 156 | 0.53 | 0.593 |
| Al | 8.095×9.051×51.203 | 156 | 0.49 | 0.532 |
| Au | 8.340×9.324×52.295 | 156 | 0.45 | 0.520 |
| Cr | 5.718×6.393×53.358 | 118 | 2.31 | 2.195 |
| Cu | 7.253×8.109×47.435 | 156 | 0.93 | 0.997 |
| Fe (BCC) | 5.689×6.361×53.166 | 118 | 1.65 | 1.892 |
| Fe (FCC) | 7.127×7.968×46.871 | 156 | 1.16 | 1.33 |
| Mo | 6.333×7.080×57.482 | 118 | 2.08 | 2.029 |
| Nb | 6.631×7.413×59.480 | 118 | 1.37 | 1.242 |
| Ni | 7.030×7.860×46.440 | 156 | 1.33 | 1.383 |
| Pd | 7.894×8.825×50.301 | 156 | 0.91 | 1.003 |
| Pt | 7.943×8.880×50.521 | 156 | 0.89 | 1.094 |
| Rh | 7.689×8.597×49.387 | 156 | 1.45 | 1.680 |
| Ta | 6.632×7.415×59.489 | 118 | 1.41 | 1.412 |
| V | 6.000×6.708×55.249 | 118 | 1.14 | 1.204 |
| W | 6.373×7.125×57.751 | 118 | 2.71 | 2.654 |

We evaluate grain boundary segregation tendency via substitution and relaxation[42,43] to assess the segregation energy $\Delta E_i^{seg}$ as[23,58,59]:

$$\Delta E_i^{seg} = E_i^{GB} - E^{Bulk} \quad (1)$$

where the energetic difference is assessed between a GB slab with a substitutional solute at a GB site ($E_i^{GB}$) compared with that of a bulk site ($E^{Bulk}$). We can approximate local site concentration in equilibrium using the classical isotherm[59,60]:

$$\frac{X_i^{GB}}{1 - X_i^{GB}} = \frac{X^C}{1 - X^C} \exp\left(\frac{-\Delta E_i^{seg}}{k_B T}\right) \quad (2)$$

where $X_i^{\text{GB}}$ and $X^{\text{C}}$ are local site concentration and bulk concentration, and $k_{\text{B}}$ and $T$ denote Boltzmann constant and temperature respectively.

To assess the embrittlement potency, we follow the method described in Ref.[40] which explicitly considers two possible fracture paths, namely, those with site 1 (core site of grain boundary) detaching to one side of the facture or the other. We define embrittlement potency ($\Delta E_i^{\text{emb}}$) as[40,61]:

$$\Delta E_i^{\text{emb}} = \Delta E_i^{\text{GB}} - \Delta E_i^{\text{FS}} = \left(E_{\text{sol},i}^{\text{GB}} - E_{\text{pure}}^{\text{GB}}\right) - \left(E_{\text{sol},i}^{\text{FS}} - E_{\text{pure}}^{\text{FS}}\right) \tag{3}$$

where the superscript FS denotes free surface of the fracture, and 'sol' indicates the surface or GB with a solute.

We first show a validation by comparing the present approach to an all-electron density functional theory (DFT) dataset using GB supercells from Ref.[23] for Al-based alloys in Fig 2a; there is very strong agreement for all solutes available within a reasonable error (mean absolute error or MAE of ~0.07 eV). We also plot a validation with the DFT dataset[23] in Fig 2b for the embrittlement potencies, which spans a larger range and thus features a somewhat higher MAE. The general agreement still holds; for both segregation and the change in GB cohesive energy, the present approach based on ML interatomic potentials provides a very reasonable accuracy in a substantial tradeoff for time, vis-à-vis full DFT calculations.

Together, the two validated energies in Table 1 and Fig. 2 provide the basis for a quantitative evaluation of embrittling potency for many alloys. This is developed for BCC Fe-based binary systems as a single example in Fig 2. The panel denotes each grain boundary site we consider. Note that site 2, 3 and 4 have degeneracy of two per site 1. We show in Fig. 3 the segregation energy and embrittlement potencies of all sites in bcc Fe Σ5[001](210). For the problem of embrittlement, it is immediately a concern to see drastic segregation of known embrittlers such as S, P and Sn, which is then confirmed in Fig. 3b. Fig. 3b also agrees well with the general trend from density functional theory calculations; first, Mo shows strengthening behavior as opposed to Pd[6]. Pb and Bi are also found to embrittle grain boundary significantly[62]. Ref.[7] also suggests that Cu reduces grain boundary cohesion, while V, Cr, Mn, Mo and W increase it. We note that while we include several elements that can be interstitial such as S and P in steels, we assume only substitutional site occupation in the high-throughput calculations in this work. Data for the rest of the solvents can be found in the supplemental material.

We now turn to the quantitative combination of segregation and embrittlement. Segregation can only lead to embrittlement if it occurs, so the segregation energy of each site is used to calculate the probability that a given site is occupied in equilibrium. This probability is equal to the average site concentration calculated with a standard isotherm expression (which accounts for configurational entropy working against the segregation energy itself):

$$X_i^{\text{GB}} = \sum F_i \left[1 + \frac{1 - X^{\text{C}}}{X^{\text{C}}} \exp\left(\frac{\Delta E_i^{\text{seg}}}{k_{\text{B}} T}\right)\right]^{-1} \tag{4}$$

where $F_i$ is the site fraction of site type '$i$' in the grain boundary. The total effect of all sites, weighted by their occupation, is required to assess the net effect on embrittlement. We use a metric called the *embrittlement product*, which selects the most embrittled fracture path for each site, and averages them weighted by their local site occupation probability ($X_i^{\text{GB}}$) and site fraction ($F_i$)[23]:

$$\langle \Delta E^{\text{emb}} \rangle = \sum_i F_i X_i^{\text{GB}} \Delta E_i^{\text{emb}} \qquad (5)$$

All four sites are then used in Eq. (4) and (5) to determine the net embrittlement products in Fig. 4 for all alloys. The general agreements still hold; many strong embrittlers also segregate (P, S, Sn, Pb and Bi). For the transition metals, only systems with higher cohesive energy such as high melting point refractory solute enhances the GB cohesion, for which we rank a chosen few in Fig. 5a as well.

Lastly, it is important to note that the segregation demonstrated here is calculated from a constant concentration, and the isotherm approach of Eq. (4) assumes the only competing state is a solid solution. To account for phase equilibrium, it is useful to truncate the allowable solid solution composition at the solubility limit, which we do here using CALPHAD thermodynamics (TCFE13 from ThermoCalc[63]):

$$X^C = \min\,(1\,\text{at.\,\%}, \text{Solubility from CALPHAD}) \qquad (6)$$

Because the solubility limits how much solute the bulk matrix can have, there is thus a corresponding limit to the segregation level at solid solution grain boundaries. In Fig. 5, we plot a bar chart showing solute with (5a) and without (5b) solubility data in TCFE13, ranked from the lowest (GB strengtheners) to the highest (GB embrittler) using $T = 1000$ K. Despite the strengthening capability of Ta vis-à-vis Nb, there is limited amount of solubility in BCC Fe at 1000 K and hence a similar behavior as Nb. Similarly for some common Fe embrittlers: S has higher embrittlement tendency, but lower solubility than P (0.0043 at.% for S vis-à-vis >1 at.% for P), which is one of the most potent embrittlers in steels. Cu, a precipitation element in steels[64], also has low solubility (0.67 at.%) and thus contributes a small amount to grain boundary embrittlement despite the lower cohesive energy of Cu vis-à-vis Fe. The embrittlement from Y also reduces significantly due to its solubility as well.

In conclusion, we have computed a large, self-consistent GB embrittlement genome for segregation and embrittlement of a model high-angle GB using a universal chemistry potential. The versatility of the model allows us to establish a genome for 15 solvents and 75 solute species (>1,000 combinations) across the periodic table. This database should find use in a variety of applications, including high strength steels and aluminum, refractory alloys, and nanocrystalline alloys.

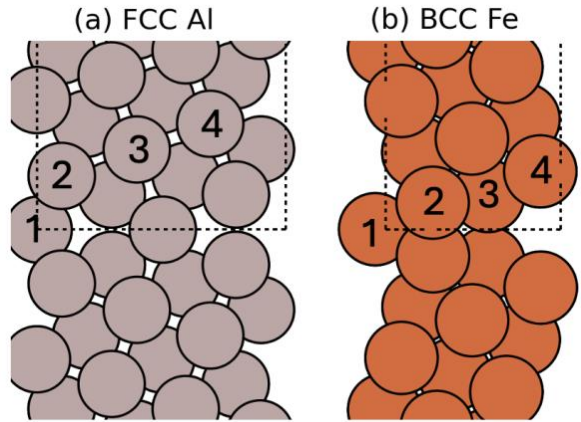

FIG. 1 Example grain boundary structure of (a) FCC Al and (b) BCC Fe Σ5[001](210). Sites 1-4 are used to calculate segregation energy and embrittlement potencies.

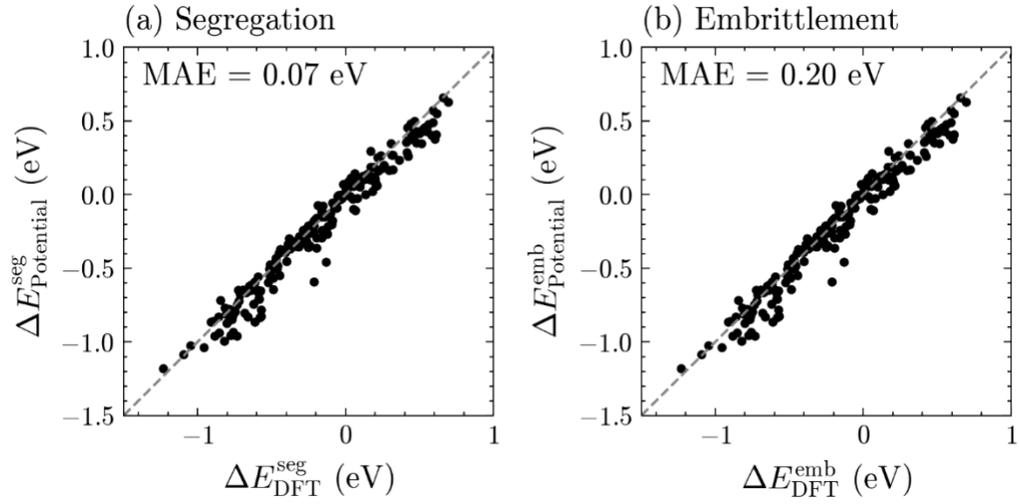

FIG. 2 Validation of the potential using DFT Σ5[001](210) grain boundary data from Ref.[23]. The mean absolute errors (MAE) are calculated for all alloys reported in Ref.[23].

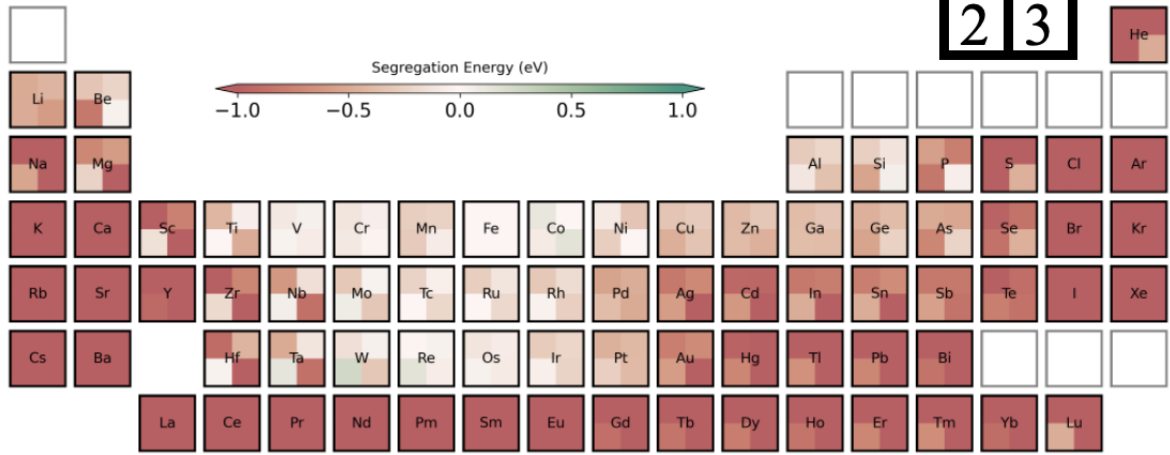

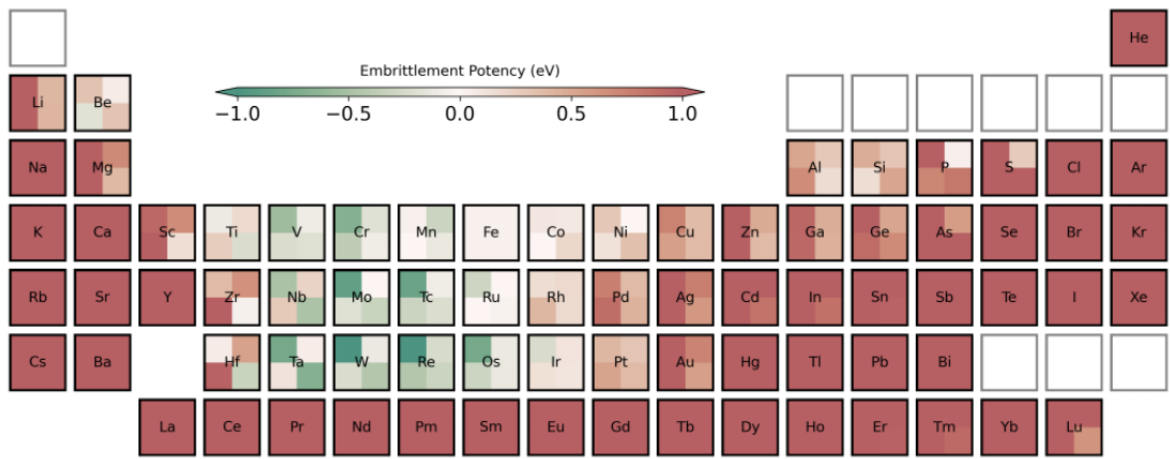

FIG. 3 (a) Segregation energies and (b) embrittlement potencies of Fe-based alloys calculated from 4 individual grain boundary sites. The most embrittle plane is presented in (b) from the two fracture planes calculated.

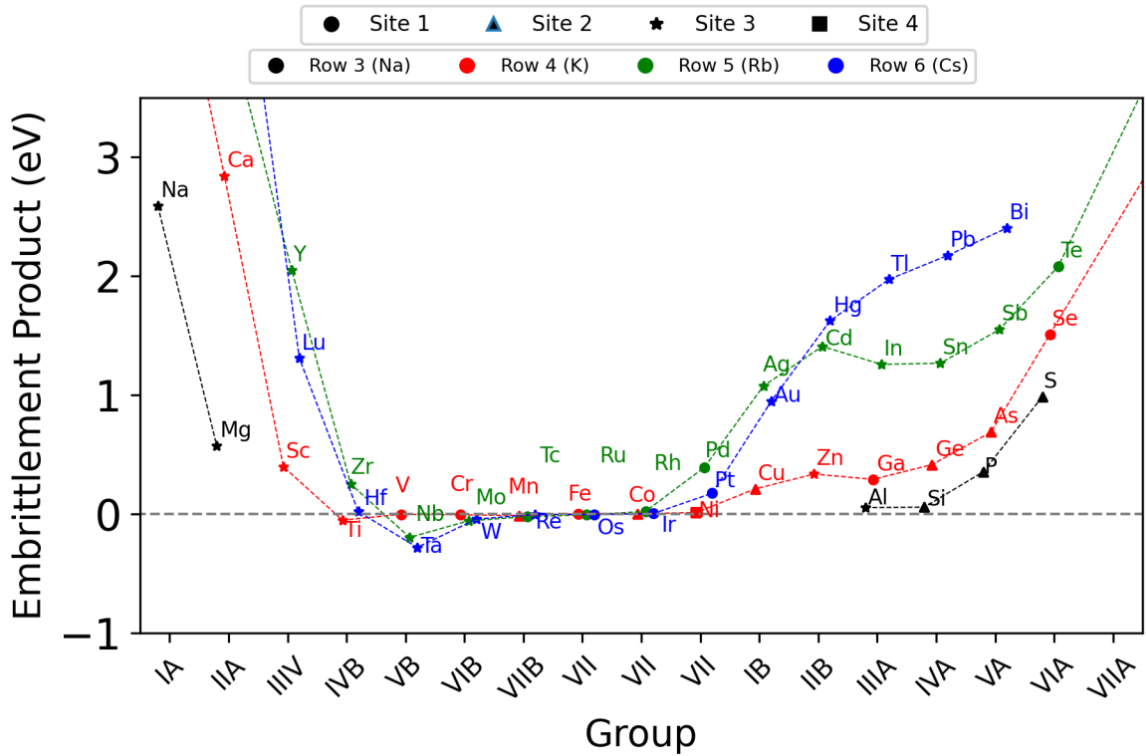

FIG. 4 Periodic summary plot of embrittlement products for Fe-based binary alloys calculated at 1000 K and 1 at. % bulk concentration.

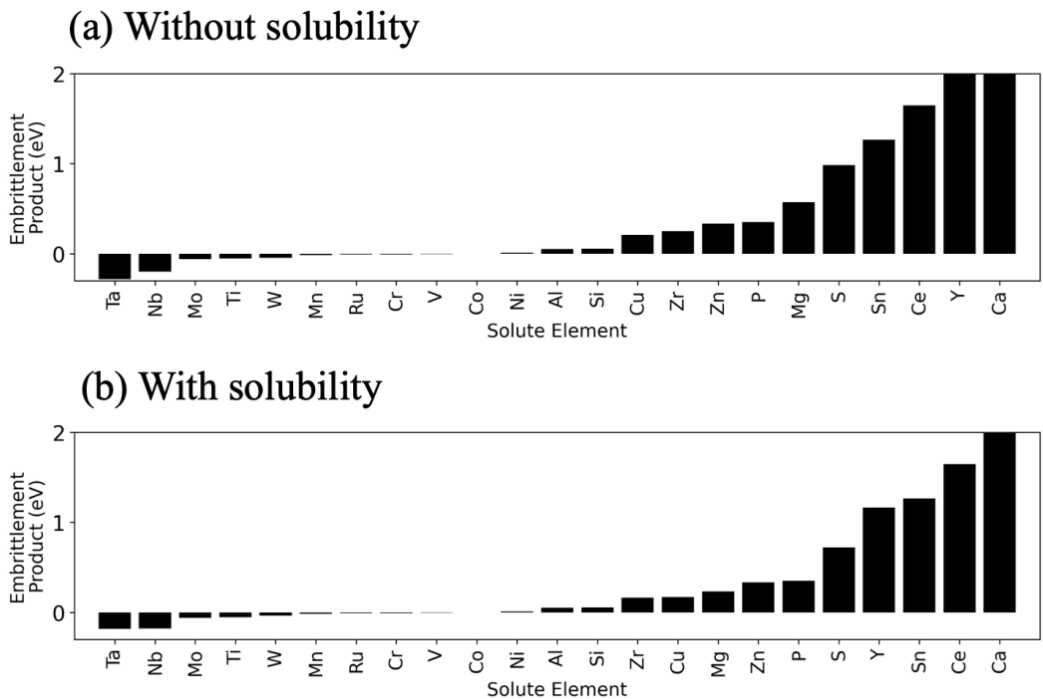

FIG. 5 Embrittlement product both without (a) and with (b) solubility data from TCFE13 at 1086 K. The bulk concentration is set at the minimum between 1 at.% or the solubility limit.

## Acknowledgements

The work is supported by Office of Naval Research (ONR) under the grant N000142312004. CAS and NT acknowledge the US Department of Energy award No. DE-SC0020180 for work on high-throughput GB genome database development. The authors acknowledge support from Constellium. The authors would like to acknowledge MIT ORCD, MIT Engaging and MIT Supercloud[65] for the HPC resources used in this work.

## Author Declarations

### Conflict of Interest

The authors have no conflicts to disclose.

### Author Contributions

**Nutth Tuchinda:** Conceptualization, Data curation, Formal analysis, Investigation, Methodology, Software, Validation, Visualization, Writing – original draft, Writing – review & editing. **Gregory B. Olson**: Conceptualization, Funding acquisition, Project administration, Resources, Supervision, Validation, Writing – original draft, Writing – review & editing. **Christopher A. Schuh**: Conceptualization, Funding acquisition, Project administration, Resources, Supervision, Validation, Writing – original draft, Writing – review & editing

## Data Availability

The data that supports the findings of this study are available within the article (and its supplementary material).

# Grain Boundary Embrittlement Genome for Substitutional Cubic Alloys

Nutth Tuchinda[a,*], Gregory B. Olson[a], Christopher A. Schuh[a,b]

[a]Department of Materials Science and Engineering, Massachusetts Institute of Technology, 77 Massachusetts Avenue, Cambridge, MA, 02139, USA

[b]Department of Materials Science and Engineering, Northwestern University, Clark Street 633, Evanston, IL, 60208, USA

*Correspondence to nutthtu@mit.edu

**Supplemental Material**

1. Summary plots for segregation energies of Σ5[001](210) grain boundary

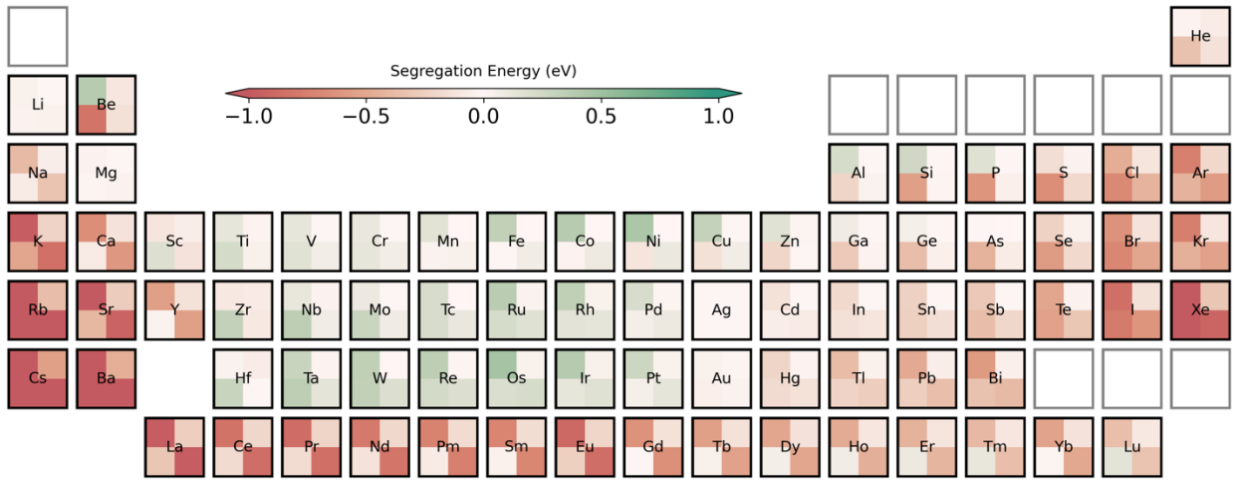

FIG. S1 Segregation energy of Ag-based systems for all 4 sites in Σ5[001](210) grain boundary. The top-left, bottom-left, bottom-right and top-right subpanel represent site 1-4 respectively (counterclockwise).

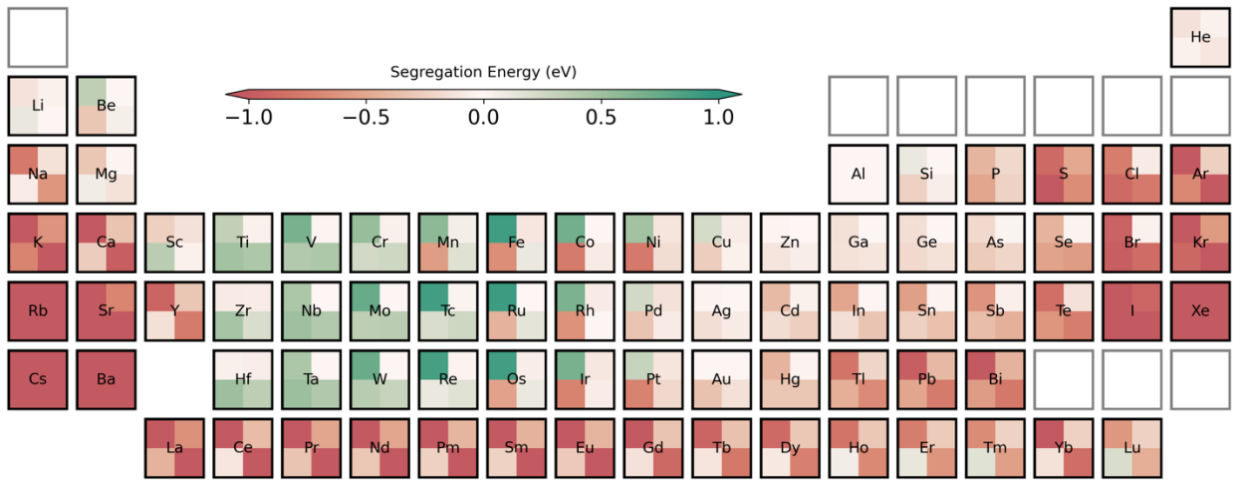

FIG. S2 Segregation energy of Al-based systems for all 4 sites in Σ5[001](210) grain boundary. The top-left, bottom-left, bottom-right and top-right subpanel represent site 1-4 respectively (counterclockwise).

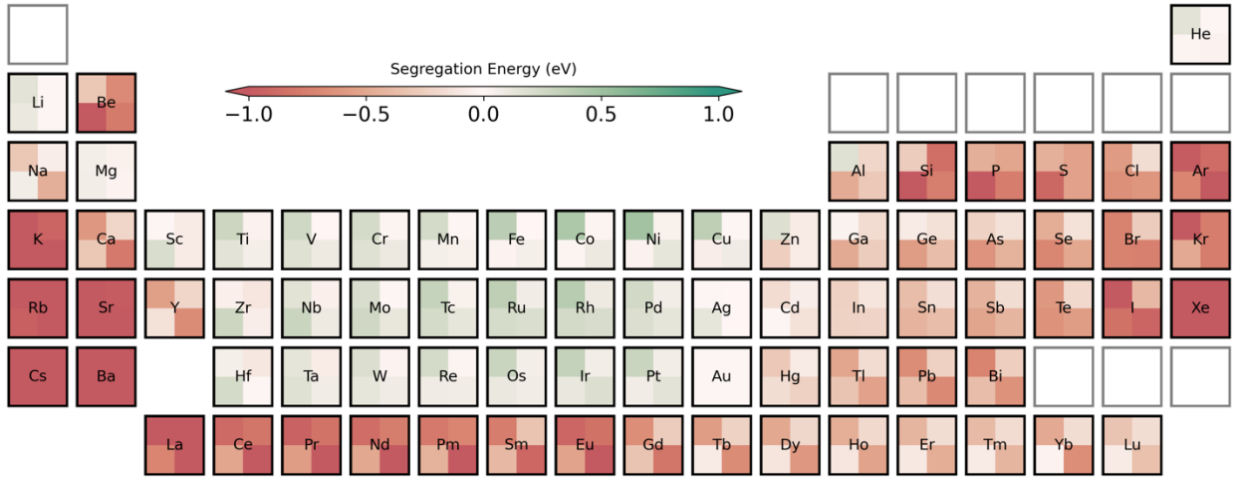

FIG. S3 Segregation energy of Au-based systems for all 4 sites in Σ5[001](210) grain boundary. The top-left, bottom-left, bottom-right and top-right subpanel represent site 1-4 respectively (counterclockwise).

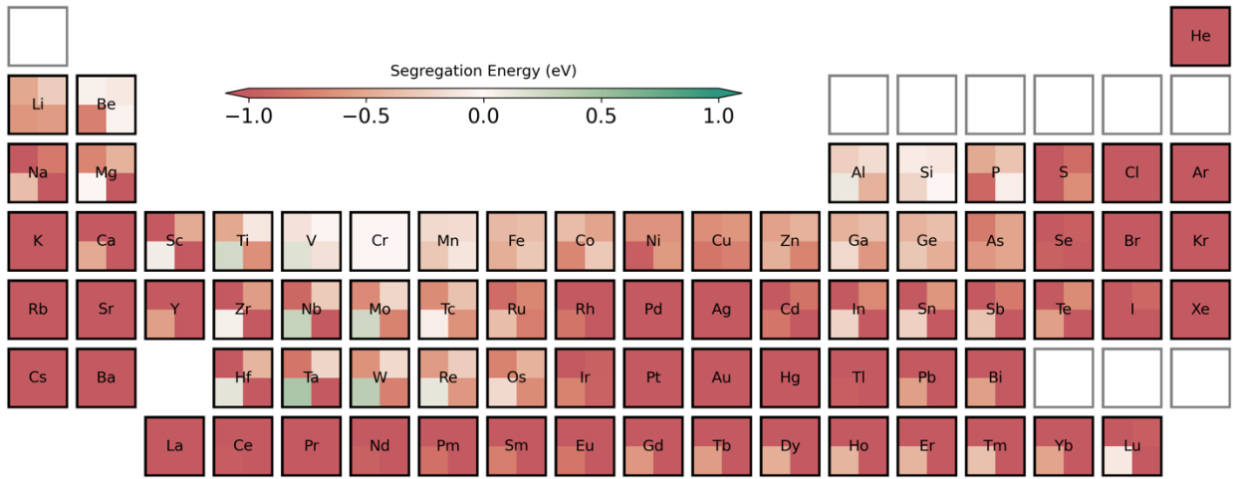

FIG. S4 Segregation energy of Cr-based systems for all 4 sites in Σ5[001](210) grain boundary. The top-left, bottom-left, bottom-right and top-right subpanel represent site 1-4 respectively (counterclockwise).

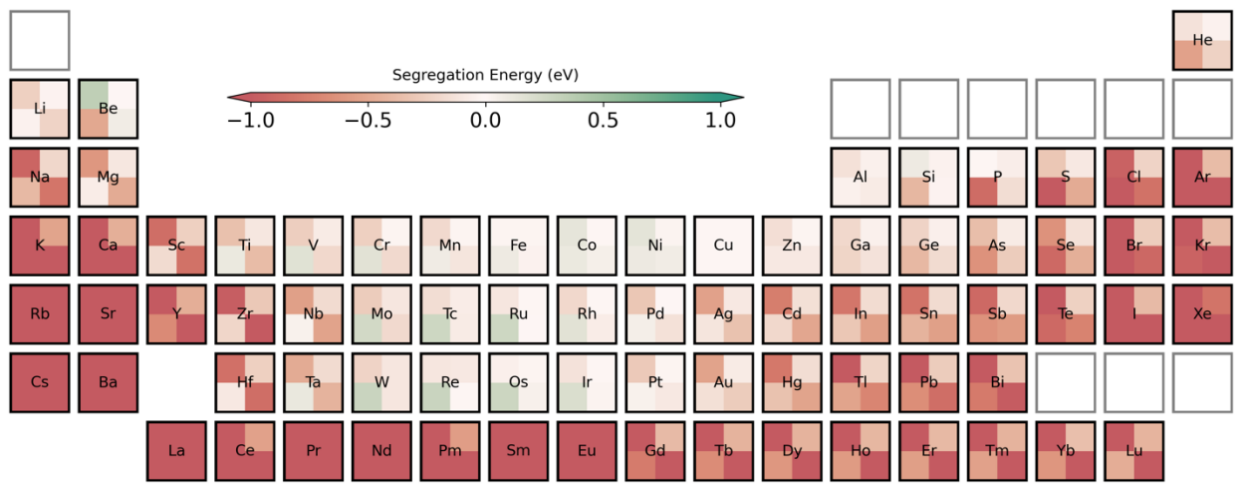

FIG. S5 Segregation energy of Cu-based systems for all 4 sites in Σ5[001](210) grain boundary. The top-left, bottom-left, bottom-right and top-right subpanel represent site 1-4 respectively (counterclockwise).

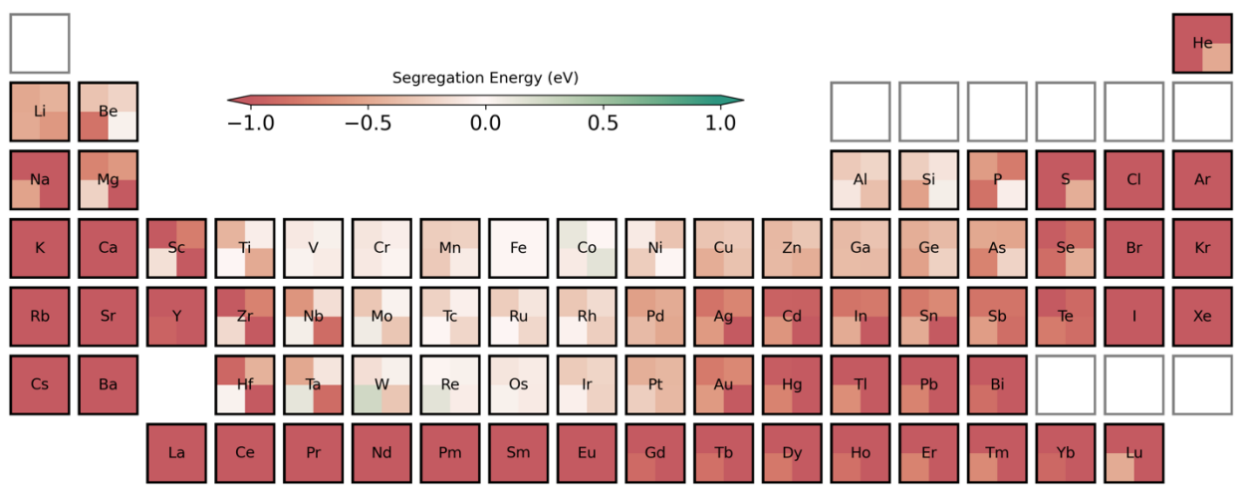

FIG. S6 Segregation energy of Fe-based systems (BCC) for all 4 sites in Σ5[001](210) grain boundary. The top-left, bottom-left, bottom-right and top-right subpanel represent site 1-4 respectively (counterclockwise).

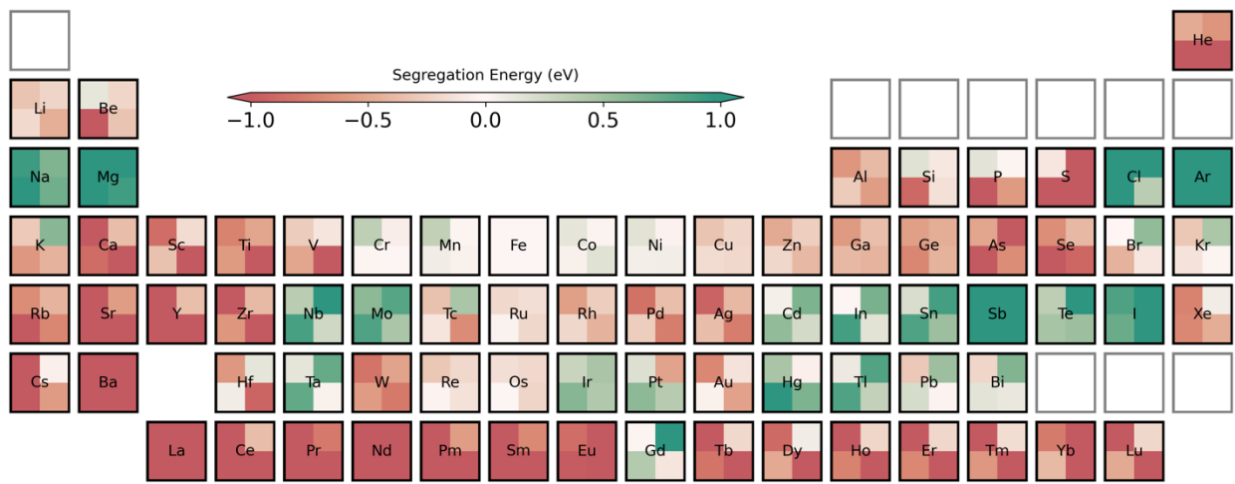

FIG. S7 Segregation energy of Fe-based systems (FCC) for all 4 sites in Σ5[001](210) grain boundary. The top-left, bottom-left, bottom-right and top-right subpanel represent site 1-4 respectively (counterclockwise).

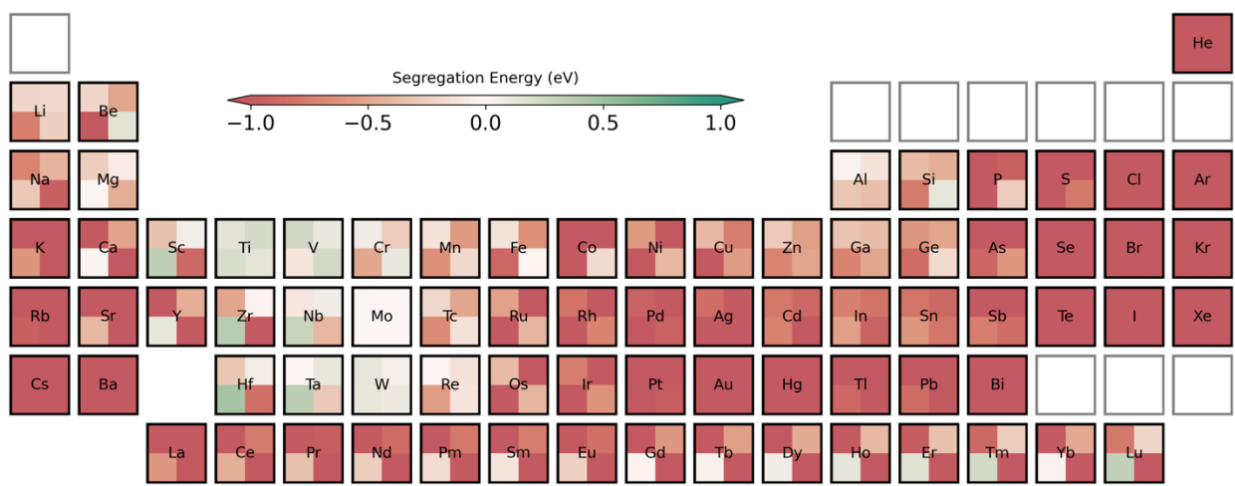

FIG. S8 Segregation energy of Mo-based systems for all 4 sites in Σ5[001](210) grain boundary. The top-left, bottom-left, bottom-right and top-right subpanel represent site 1-4 respectively (counterclockwise).

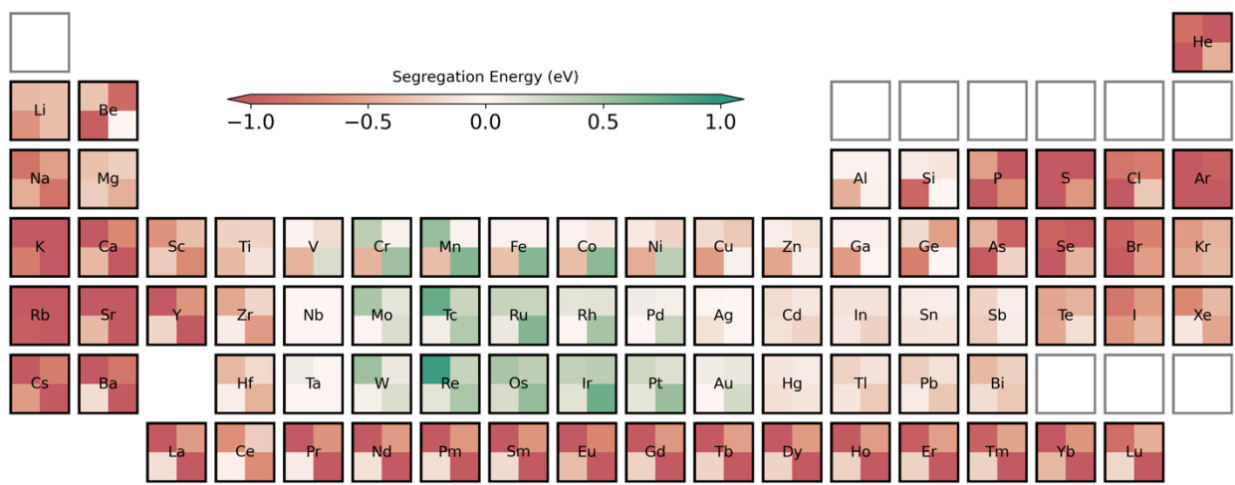

FIG. S9 Segregation energy of Nb-based systems for all 4 sites in Σ5[001](210) grain boundary. The top-left, bottom-left, bottom-right and top-right subpanel represent site 1-4 respectively (counterclockwise).

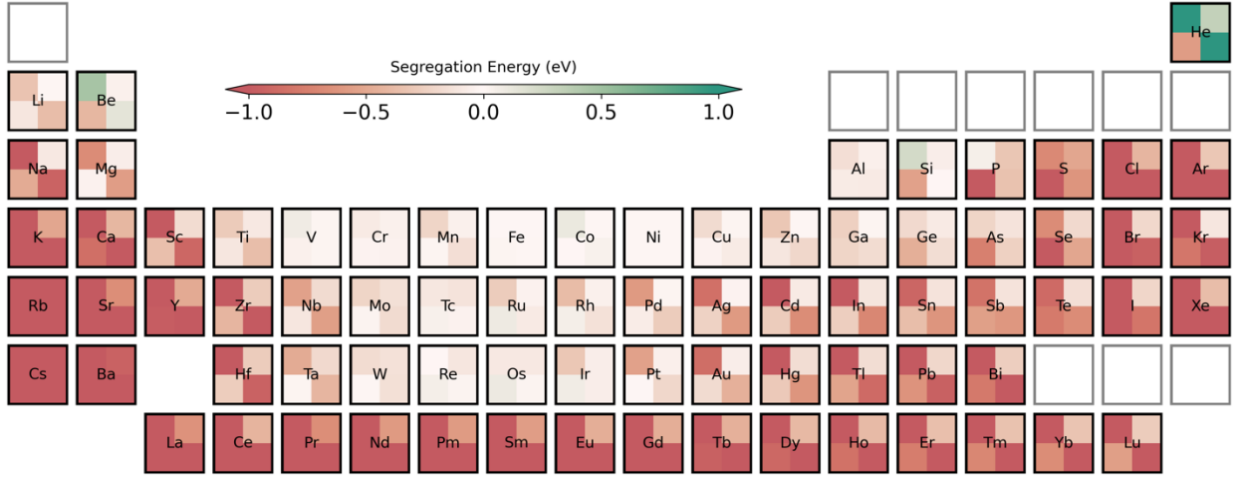

FIG. S10 Segregation energy of Ni-based systems for all 4 sites in Σ5[001](210) grain boundary. The top-left, bottom-left, bottom-right and top-right subpanel represent site 1-4 respectively (counterclockwise).

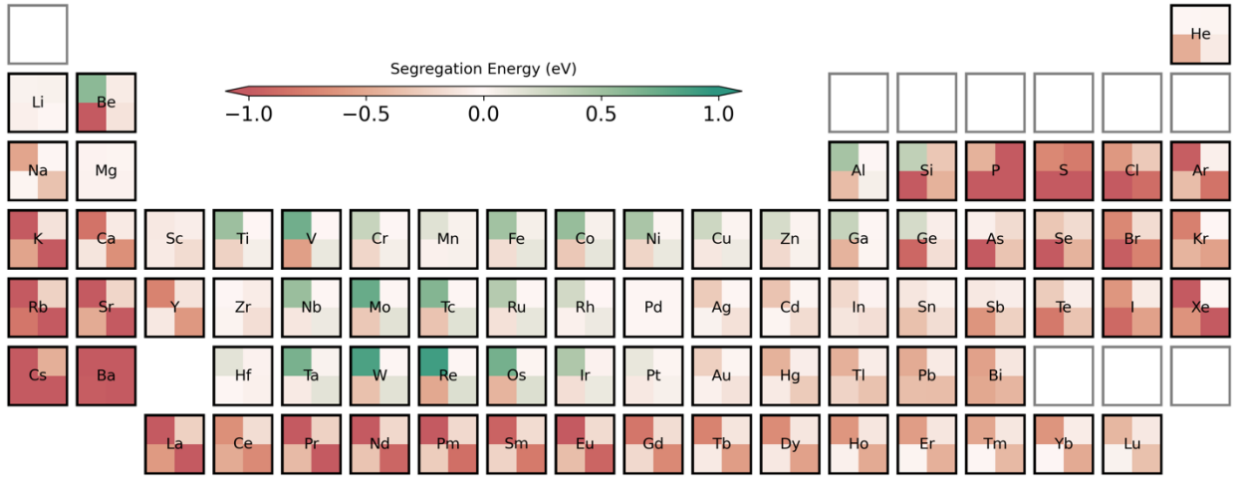

FIG. S11 Segregation energy of Pd-based systems for all 4 sites in Σ5[001](210) grain boundary. The top-left, bottom-left, bottom-right and top-right subpanel represent site 1-4 respectively (counterclockwise).

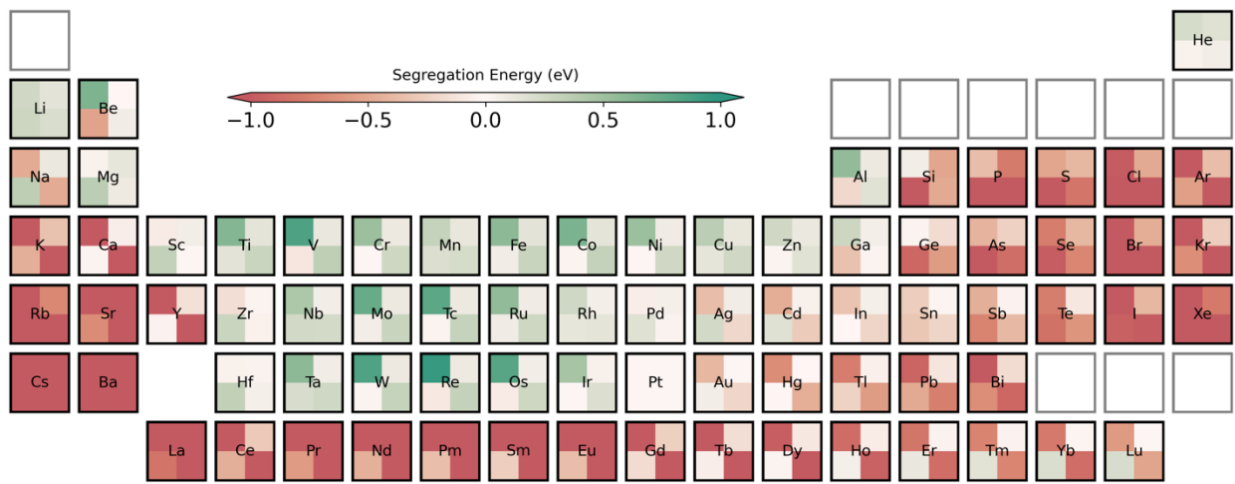

FIG. S12 Segregation energy of Pt-based systems for all 4 sites in Σ5[001](210) grain boundary. The top-left, bottom-left, bottom-right and top-right subpanel represent site 1-4 respectively (counterclockwise).

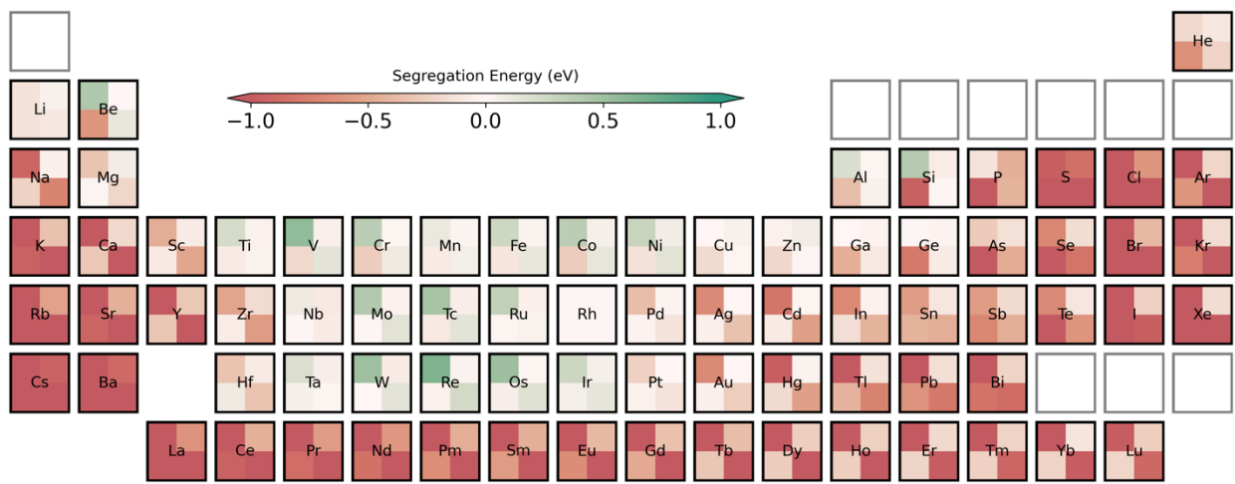

FIG. S13 Segregation energy of Rh-based systems for all 4 sites in Σ5[001](210) grain boundary. The top-left, bottom-left, bottom-right and top-right subpanel represent site 1-4 respectively (counterclockwise).

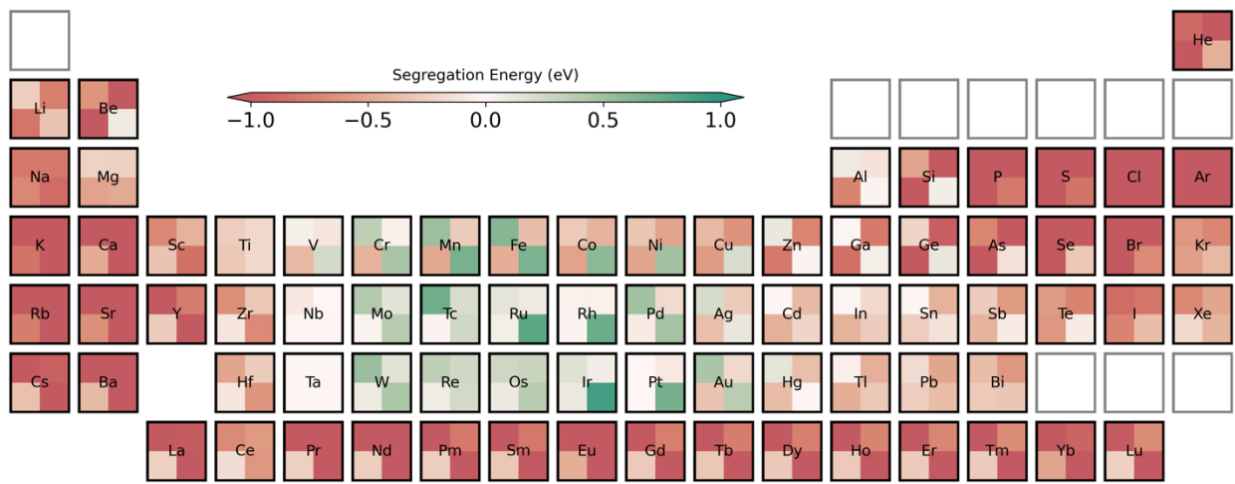

FIG. S14 Segregation energy of Ta-based systems for all 4 sites in Σ5[001](210) grain boundary. The top-left, bottom-left, bottom-right and top-right subpanel represent site 1-4 respectively (counterclockwise).

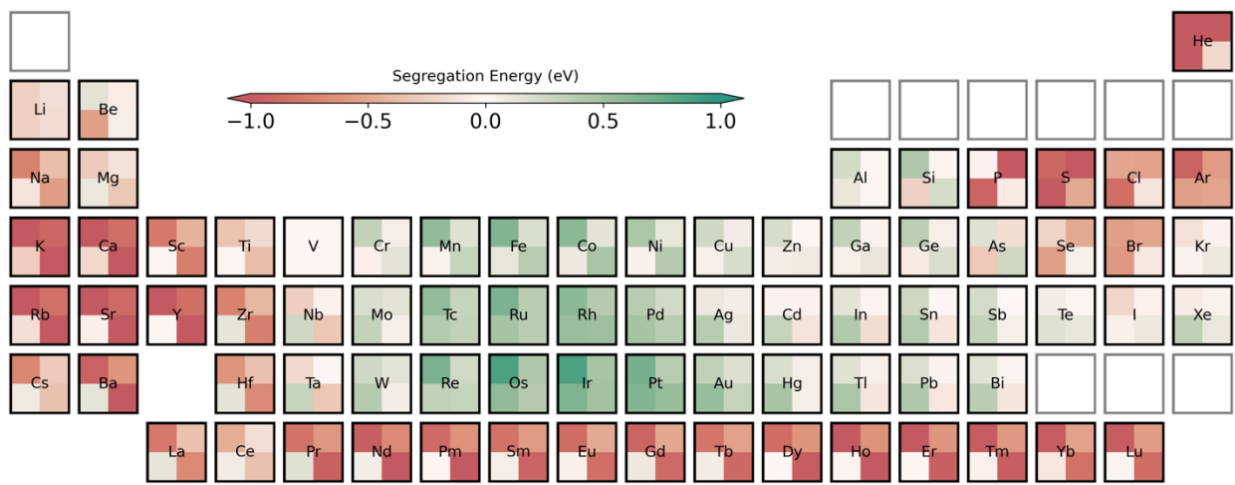

FIG. S15 Segregation energy of V-based systems for all 4 sites in Σ5[001](210) grain boundary. The top-left, bottom-left, bottom-right and top-right subpanel represent site 1-4 respectively (counterclockwise).

FIG. S16 Segregation energy of W-based systems for all 4 sites in Σ5[001](210) grain boundary. The top-left, bottom-left, bottom-right and top-right subpanel represent site 1-4 respectively (counterclockwise).

## 2. Summary plots for embrittlement potencies of Σ5[001](210) grain boundary

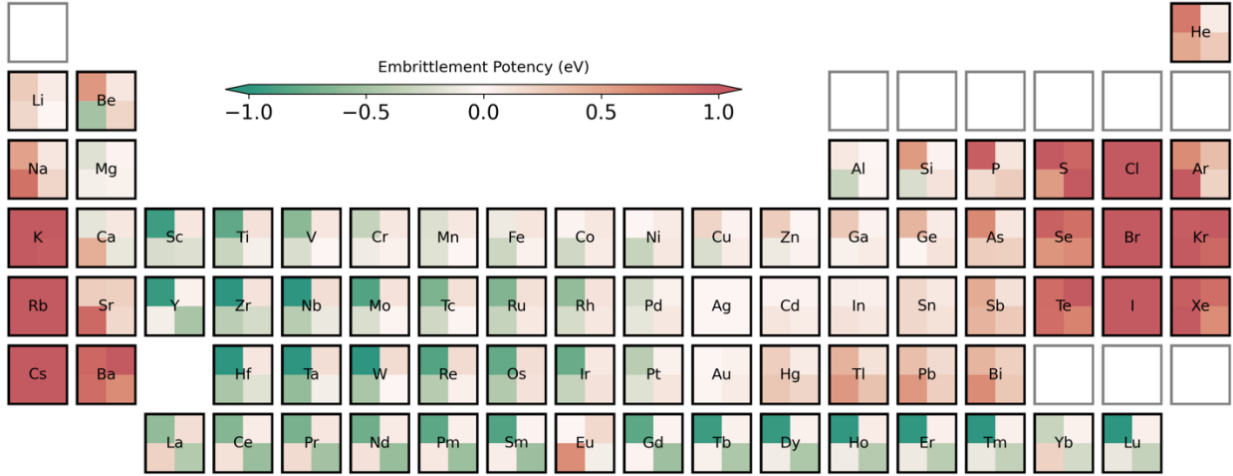

FIG. S17 Embrittlement potency of Ag-based systems for all 4 sites in Σ5[001](210) grain boundary. The top-left, bottom-left, bottom-right and top-right subpanel represent site 1-4 respectively (counterclockwise). The highest embrittlement potency is chosen for site 2-4 from the fracture planes calculated.

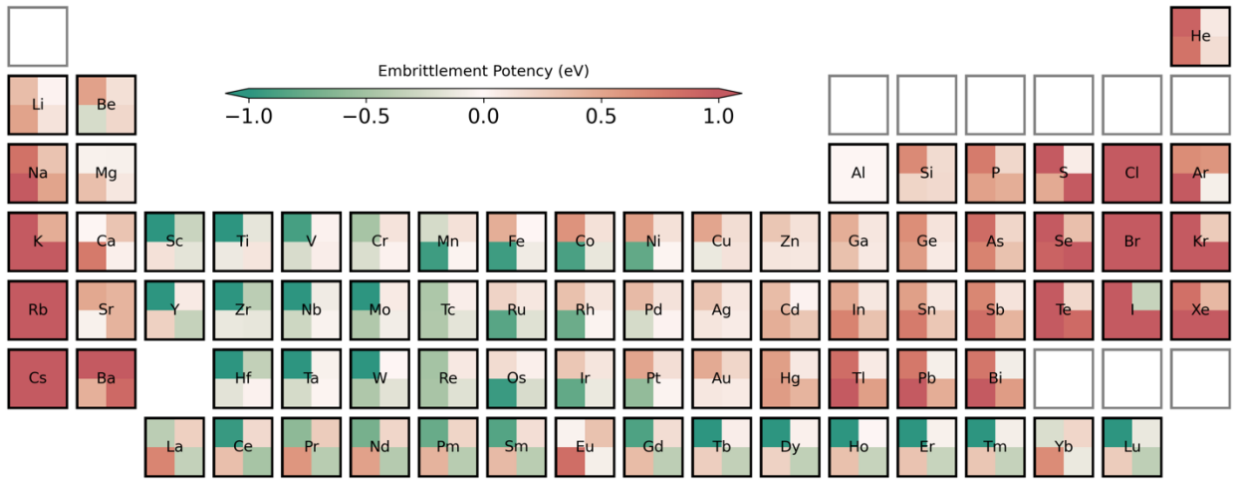

FIG. S18 Embrittlement potency of Al-based systems for all 4 sites in Σ5[001](210) grain boundary. The top-left, bottom-left, bottom-right and top-right subpanel represent site 1-4 respectively (counterclockwise). The highest embrittlement potency is chosen for site 2-4 from the fracture planes calculated.

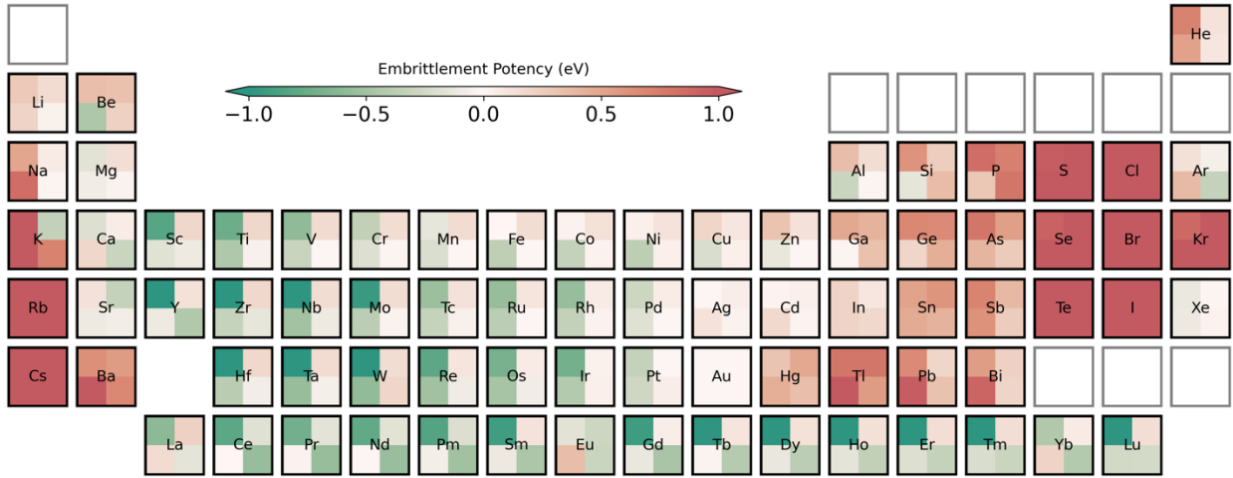

FIG. S19 Embrittlement potency of Au-based systems for all 4 sites in Σ5[001](210) grain boundary. The top-left, bottom-left, bottom-right and top-right subpanel represent site 1-4 respectively (counterclockwise). The highest embrittlement potency is chosen for site 2-4 from the fracture planes calculated.

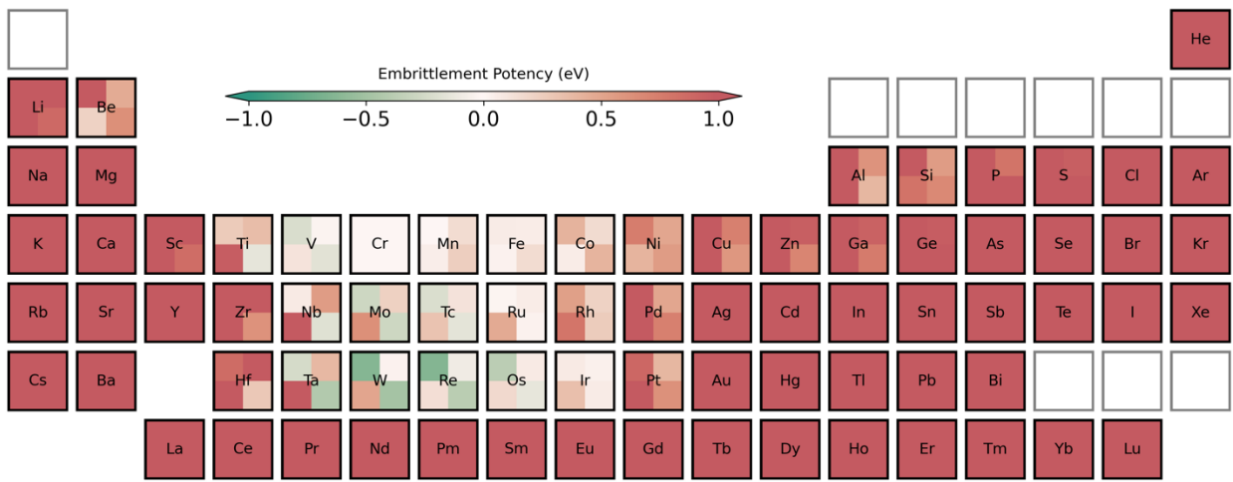

FIG. S20 Embrittlement potency of Cr-based systems for all 4 sites in Σ5[001](210) grain boundary. The top-left, bottom-left, bottom-right and top-right subpanel represent site 1-4 respectively (counterclockwise). The highest embrittlement potency is chosen for site 2-4 from the fracture planes calculated.

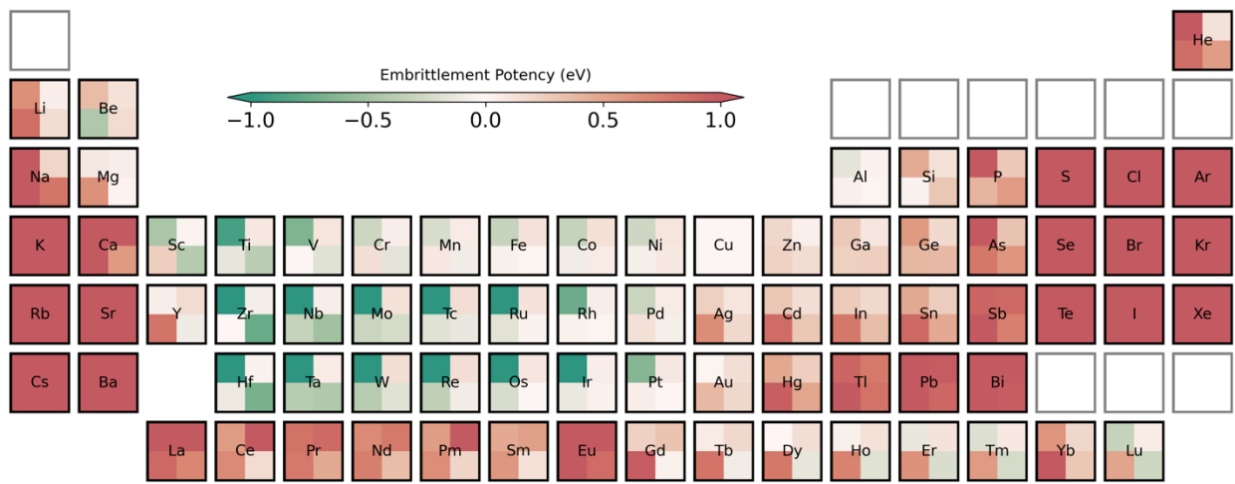

FIG. S21 Embrittlement potency of Cu-based systems for all 4 sites in Σ5[001](210) grain boundary. The top-left, bottom-left, bottom-right and top-right subpanel represent site 1-4 respectively (counterclockwise). The highest embrittlement potency is chosen for site 2-4 from the fracture planes calculated.

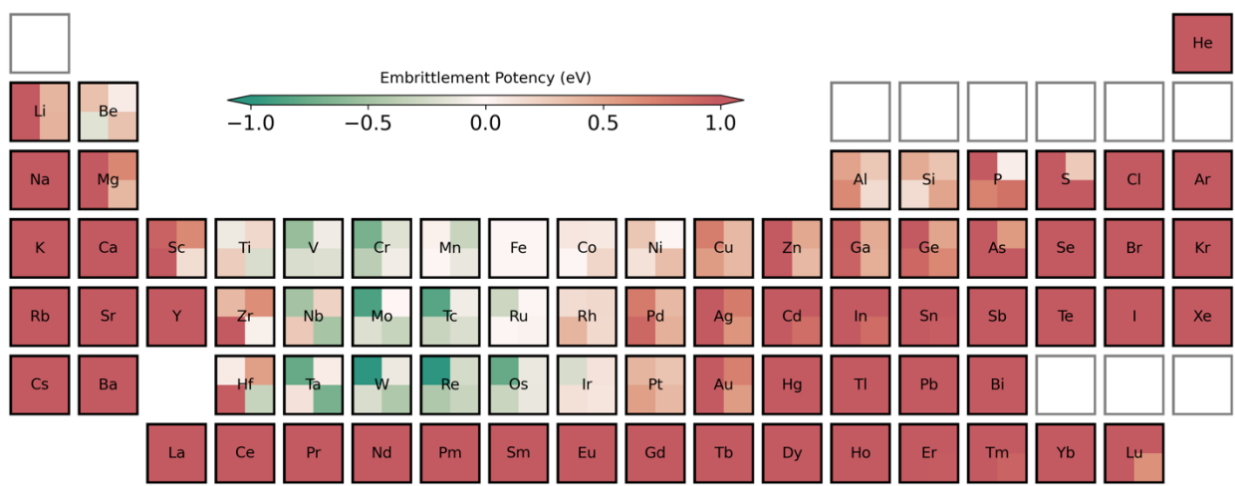

FIG. S22 Embrittlement potency of Fe-based systems (BCC) for all 4 sites in Σ5[001](210) grain boundary. The top-left, bottom-left, bottom-right and top-right subpanel represent site 1-4 respectively (counterclockwise). The highest embrittlement potency is chosen for site 2-4 from the fracture planes calculated.

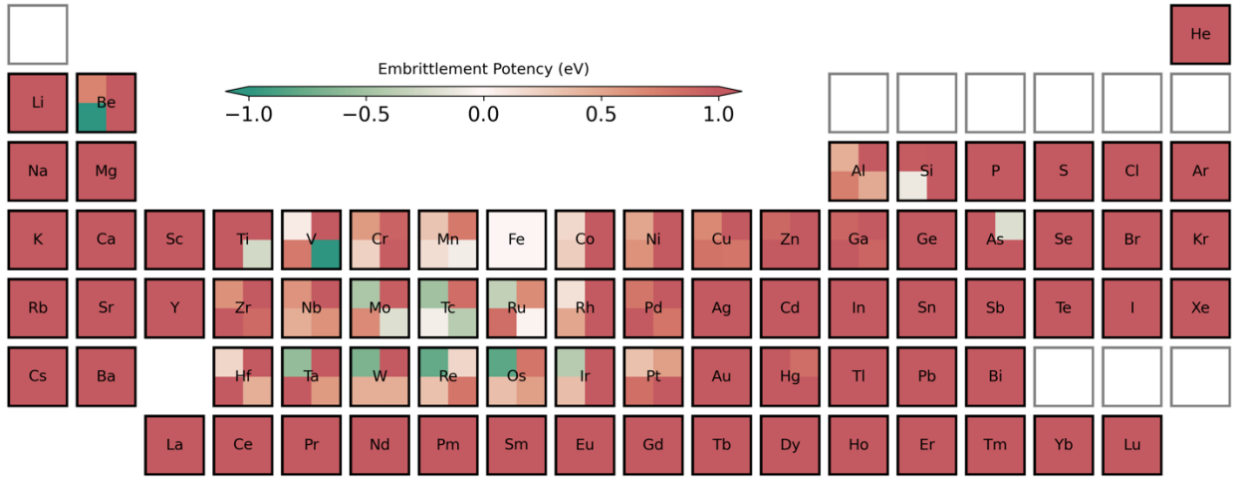

FIG. S23 Embrittlement potency of Fe-based systems (FCC) for all 4 sites in Σ5[001](210) grain boundary. The top-left, bottom-left, bottom-right and top-right subpanel represent site 1-4 respectively (counterclockwise). The highest embrittlement potency is chosen for site 2-4 from the fracture planes calculated.

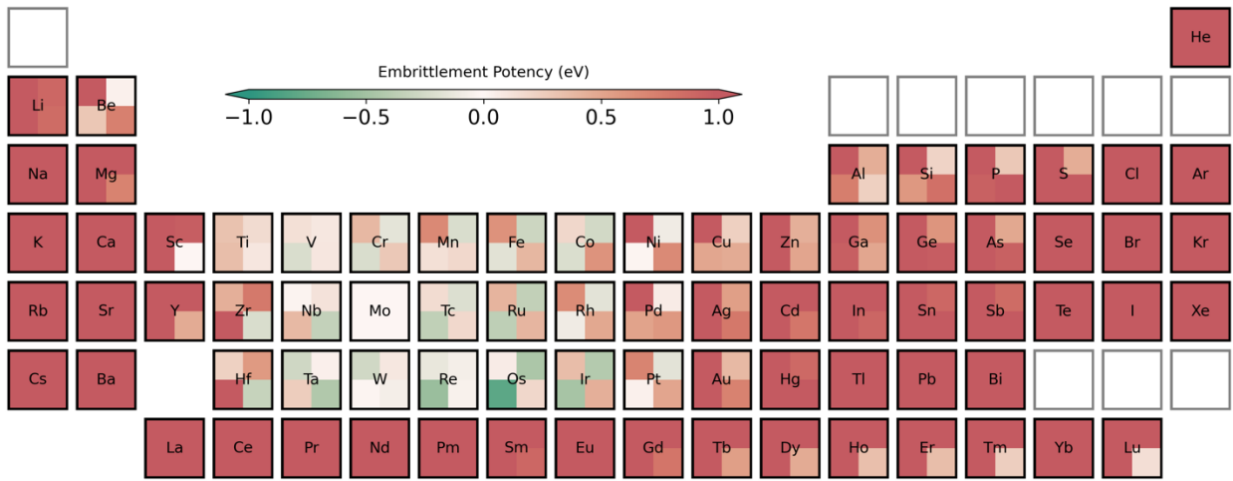

FIG. S24 Embrittlement potency of Mo-based systems for all 4 sites in Σ5[001](210) grain boundary. The top-left, bottom-left, bottom-right and top-right subpanel represent site 1-4 respectively (counterclockwise). The highest embrittlement potency is chosen for site 2-4 from the fracture planes calculated.

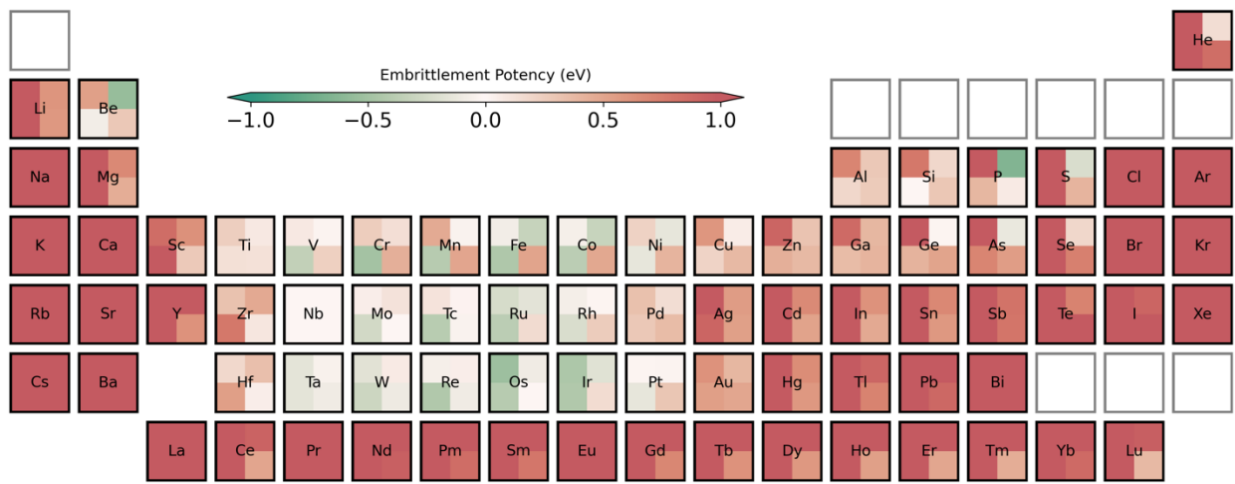

FIG. S25 Embrittlement potency of Nb-based systems for all 4 sites in Σ5[001](210) grain boundary. The top-left, bottom-left, bottom-right and top-right subpanel represent site 1-4 respectively (counterclockwise). The highest embrittlement potency is chosen for site 2-4 from the fracture planes calculated.

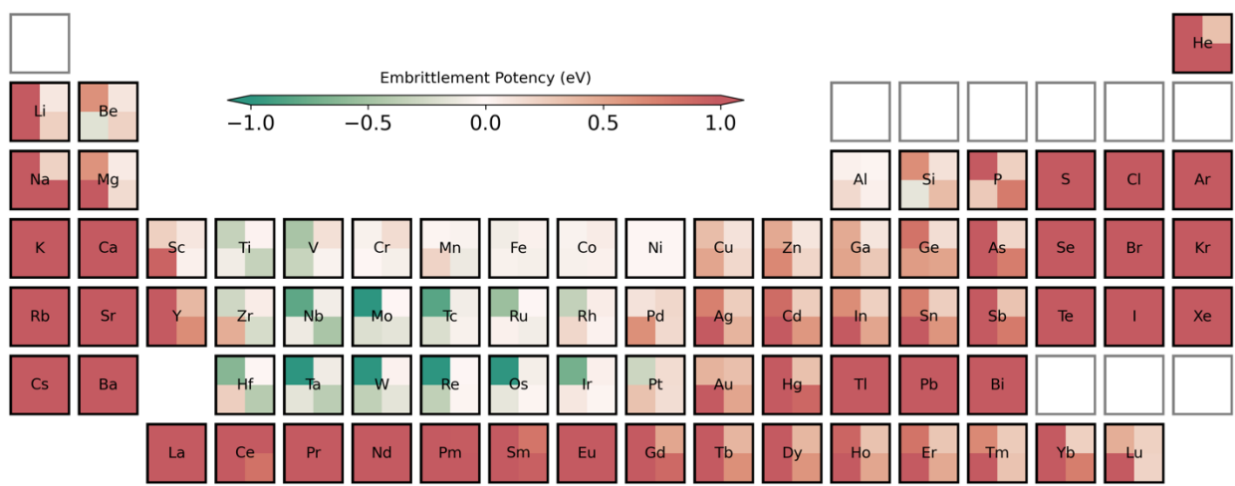

FIG. S26 Embrittlement potency of Ni-based systems for all 4 sites in Σ5[001](210) grain boundary. The top-left, bottom-left, bottom-right and top-right subpanel represent site 1-4 respectively (counterclockwise). The highest embrittlement potency is chosen for site 2-4 from the fracture planes calculated.

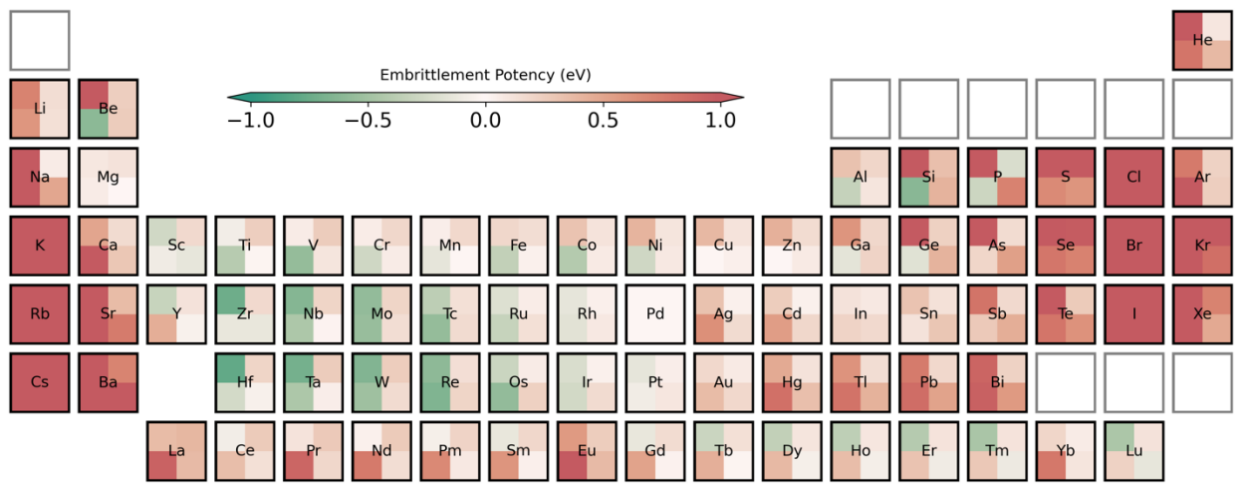

FIG. S27 Embrittlement potency of Pd-based systems for all 4 sites in Σ5[001](210) grain boundary. The top-left, bottom-left, bottom-right and top-right subpanel represent site 1-4 respectively (counterclockwise). The highest embrittlement potency is chosen for site 2-4 from the fracture planes calculated.

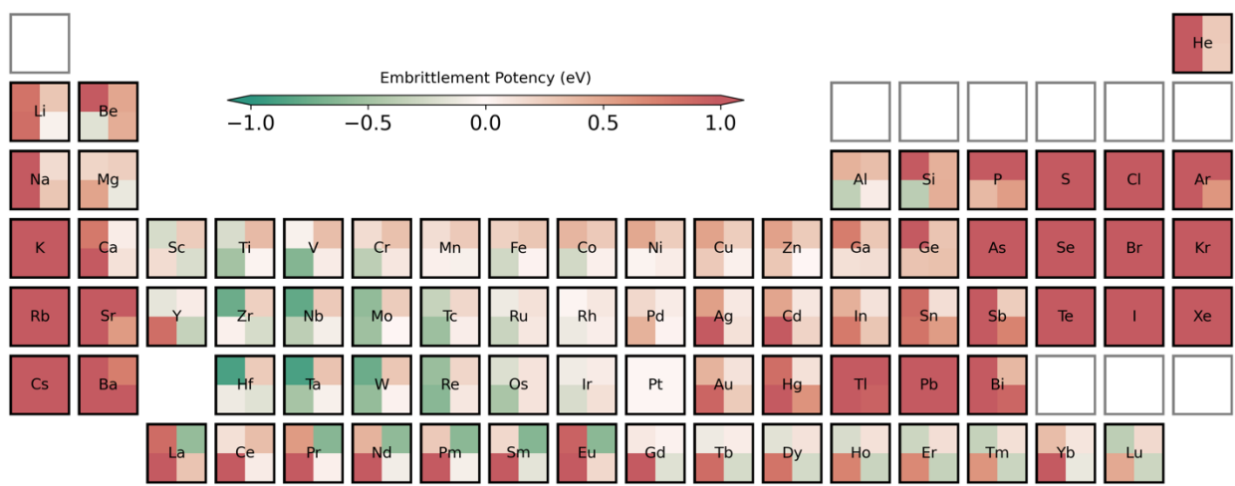

FIG. S28 Embrittlement potency of Pt-based systems for all 4 sites in Σ5[001](210) grain boundary. The top-left, bottom-left, bottom-right and top-right subpanel represent site 1-4 respectively (counterclockwise). The highest embrittlement potency is chosen for site 2-4 from the fracture planes calculated.

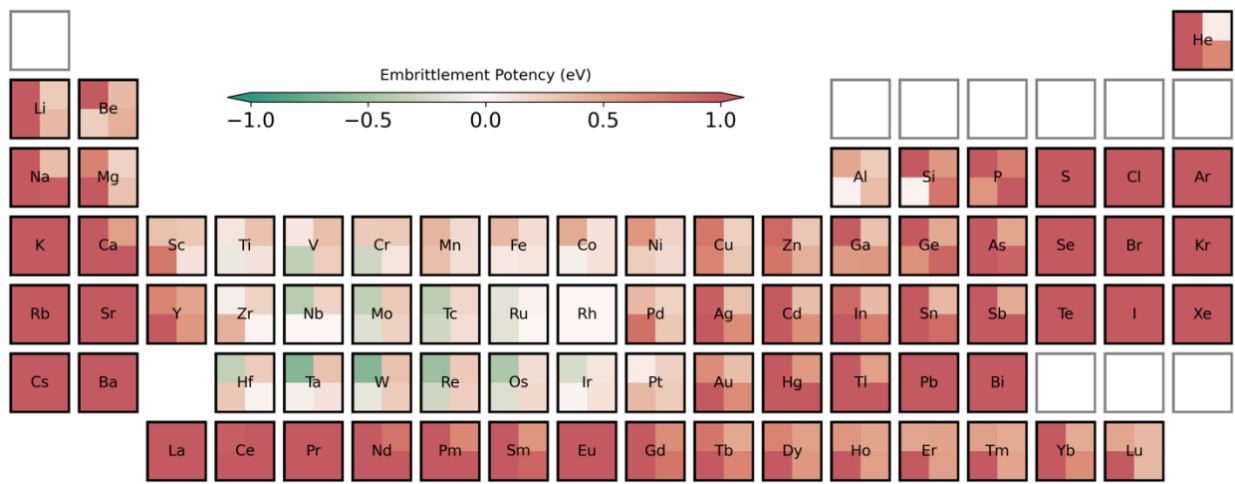

FIG. S29 Embrittlement potency of Rh-based systems for all 4 sites in Σ5[001](210) grain boundary. The top-left, bottom-left, bottom-right and top-right subpanel represent site 1-4 respectively (counterclockwise). The highest embrittlement potency is chosen for site 2-4 from the fracture planes calculated.

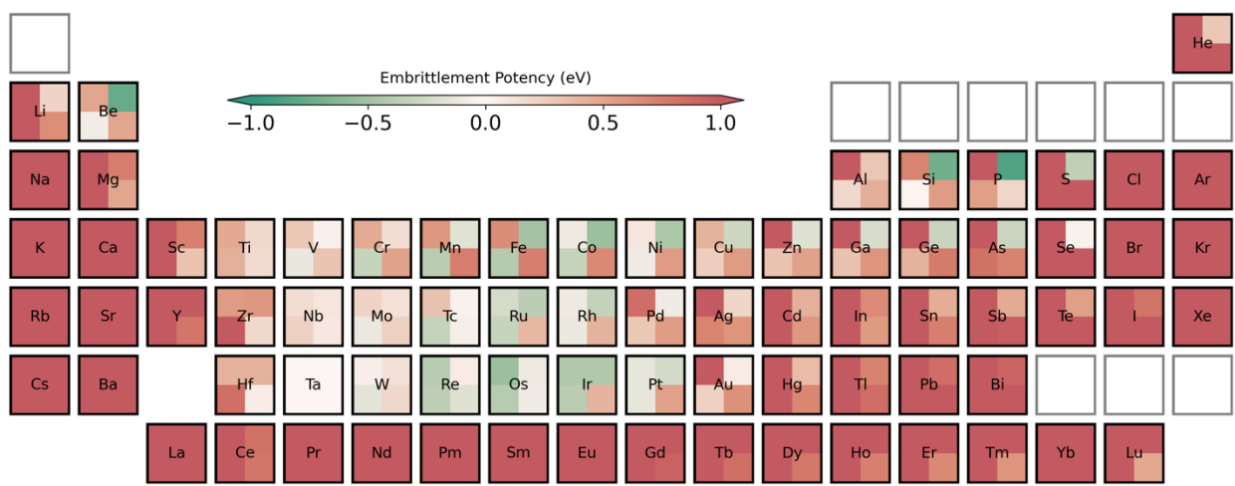

FIG. S30 Embrittlement potency of Ta-based systems for all 4 sites in Σ5[001](210) grain boundary. The top-left, bottom-left, bottom-right and top-right subpanel represent site 1-4 respectively (counterclockwise). The highest embrittlement potency is chosen for site 2-4 from the fracture planes calculated.

FIG. S31 Embrittlement potency of V-based systems for all 4 sites in Σ5[001](210) grain boundary. The top-left, bottom-left, bottom-right and top-right subpanel represent site 1-4 respectively (counterclockwise). The highest embrittlement potency is chosen for site 2-4 from the fracture planes calculated.

FIG. S32 Embrittlement potency of W-based systems for all 4 sites in Σ5[001](210) grain boundary. The top-left, bottom-left, bottom-right and top-right subpanel represent site 1-4 respectively (counterclockwise). The highest embrittlement potency is chosen for site 2-4 from the fracture planes calculated.